8/15/2005

# Inferring interactions from combinatorial protein libraries

Jeffrey B. Endelman[1], Jesse D. Bloom[2], Christopher R. Otey[3], Marco Landwehr[2] and Frances H. Arnold[1,2,3,4]

[1]Bioengineering, [2]Division of Chemistry and Chemical Engineering, [3]Biochemistry and Molecular Biophysics, California Institute of Technology, Mail Code 210-41, Pasadena, CA 91125

**Manuscript information:** 21 pages, 2 figures

[4]To whom correspondence should be addressed:
Frances H. Arnold
Division of Chemistry and Chemical Engineering
California Institute of Technology, Mail Code 210-41
Pasadena, CA 91125
Tel: (626) 395-4162
Fax: (626) 568-8743
E-mail: frances@cheme.caltech.edu

## Abstract

Proteins created by combinatorial methods *in vitro* are an important source of information for understanding sequence-structure-function relationships. Alignments of folded proteins from combinatorial libraries can be analyzed using methods developed for naturally occurring proteins, but this neglects the information contained in the unfolded sequences of the library. We introduce two algorithms, logistic regression and excess information analysis, that use both the folded and unfolded sequences and compare them against contingency table and statistical coupling analysis, which only use the former. The test set for this benchmark study is a library of fictitious proteins that fold according to a hypothetical energy model. Of the four methods studied, only logistic regression is able to correctly recapitulate the energy model from the sequence alignment. The other algorithms predict spurious interactions between alignment positions with strong but individual influences on protein stability. When present in the same protein, stabilizing amino acids tend to lower the energy below the threshold needed for folding. As a result, their frequencies in the alignment can be correlated even if the positions do not interact. We believe any algorithm that neglects the nonlinear relationship between folding and energy is susceptible to this error.

## Introduction

Multiple sequence alignments (MSAs) of naturally occurring proteins have been intensely studied to better understand how sequence determines fold and function.[1-6] Such analyses are necessarily complicated by the other evolutionary influences on each protein in its biological context. Further difficulties arise because natural MSAs represent an extremely sparse sampling of the set of all amino acid combinations possible with the aligned sequences. For example, in an alignment with 50 variable positions and two different amino acids at each position, the number of possible sequences is $2^{50} \approx 10^{15}$.

Artificial families of proteins that suffer neither of these problems can be created *in vitro*, making them attractive as a complementary source of sequence-structure-function relationships. We focus here on the class of combinatorial protein libraries shown in Figure 1A, in which specific amino acids fragments of any length, including one residue, are incorporated at a fixed number of variable positions.[7-9] The composition of these libraries is determined during the design, which facilitates optimization[10-12] of their properties as well as inexpensive, high-throughput sequencing by complementary nucleotide probes.[13] When aligning sequences from this class of libraries, each variable position is represented as a single column, regardless of how many residues are present (see Figure 1C).

For modest library sizes, it is possible to characterize much of the library by high-throughput screening methods,[14] thereby mitigating the problem of sampling encountered with natural MSAs. We recently constructed such a data set by characterizing 806 unique, combinatorially assembled cytochrome P450 heme domains for their ability to fold and bind the heme cofactor.[15] The library from which these sequences were derived was constructed by shuffling three fragments at each of eight positions, for a total of $3^8 = 6,561$ possible sequences.

The fragments were chosen from three bacterial cytochrome P450 homologs to span the complete length of the heme domain, a strategy we call site-directed recombination (see Figure 1B). Of the 806 sequences in our alignment, 465 fold and bind heme. This subset of the full MSA could be analyzed using methods developed for natural protein families, but such an approach would neglect the information contained in the remaining 341 sequences.

In this manuscript we describe two new algorithms, excess information and logistic regression, for the analysis of MSAs that contain both folded and unfolded sequences. Excess information analysis is based on the mutual information between folding and a pair of positions. This differs from (but is related to) the mutual information between two positions in an MSA of folded proteins.[6,16,17] Logistic regression is an analog of linear regression for modeling binary data (e.g., 1 = folded, 0 = unfolded) and is widely used in the medical and social sciences.[18,19]

Before applying one of these methods to our P450 data (for the detailed results of the P450 analysis, see Otey et al.[15]), we conducted a benchmark study to compare them against contingency table[6,20] and statistical coupling[2,21] analysis, two methods commonly used with natural MSAs. The results of this study are presented here. The benchmark is a combinatorial library of $3^8 = 6,561$ fictitious proteins, designed as in Figure 1, of which half are folded. Folding in this library is determined by a hypothetical energy model, in which the total energy of each sequence is the sum of 8 one-body terms, one for each fragment, and $(7 \times 8)/2 = 28$ two-body terms, one for each fragment-fragment combination. If the energy of a sequence is above an arbitrary threshold of zero, it is unfolded; otherwise it is folded. This threshold model mimics the fact that naturally occurring proteins must meet certain minimal requirements for stability to fold and function in their native environments.[22,23] Using only the MSA and with no knowledge of the energy model, each algorithm must score the relative strength of all 28 two-body interactions between the positions. Accuracy is defined as the extent to which these scores agree with the energy model.

## Methods

### Construction of the fictitious library

The fictitious library contains 6,561 sequences, representing all combinations of 3 fragments at 8 positions. Fragment *i.x* refers to fragment *x* at position *i*. The total energy of each sequence is the sum of 8 one-body terms ($\varepsilon_1$) and 28 two-body terms ($\varepsilon_2$):

$$E = \sum_{i=1}^{8} \varepsilon_1(i.x) + \sum_{i=1}^{7} \sum_{j=i+1}^{8} \varepsilon_2(i.x, j.y). \quad [1]$$

There are 3 one-body terms for every position (one per fragment) and 3×3 = 9 two-body terms for every pair of positions (one per fragment-fragment combination). Of the 8 positions and 28 pairs, 7 positions and 1 pair were selected to make energetic contributions. The parameters for these variables, listed in Tables I and II, were chosen randomly from the standard normal distribution and constrained to have zero average energy. All other energy parameters were set equal to zero. The larger the differences between parameters for a particular position or pair, the more strongly it affects folding. With a stability threshold of zero energy, roughly half the library is folded (3,272 out of 6,561 sequences).

### Contingency table analysis

Several variations on contingency table analysis have been used to detect correlated residues in natural protein families.[5,6,20] For every position pair *i-j*, we tabulated the number of times each fragment-fragment combination *i.x-j.y* was observed in the folded subset of the library. Then we calculated the number expected if the two fragments were distributed independently.[24] The chi-square statistic quantifies the significance of the differences between the observed and expected values:

$$\chi_{ij}^2 = \sum_x \sum_y \frac{[\text{Observed}(i.x\text{-}j.y) - \text{Expected}(i.x\text{-}j.y)]^2}{\text{Expected}(i.x\text{-}j.y)}. \quad [2]$$

The double sum is over all nine fragment combinations. Larger $\chi^2$ values indicate greater deviations from the hypothesis that fragments are distributed independently.

**Statistical coupling analysis**

Statistical coupling analysis, developed by Ranganathan and co-workers to measure energetic coupling in natural protein families,[2,21] was adapted for the folded subset of our combinatorial library. The statistical coupling between positions $i$ and $j$ measures the response at position $i$ when the MSA is perturbed at position $j$ ($\Delta\Delta G_{i,j}$) or vice versa ($\Delta\Delta G_{j,i}$). In general these energy vectors will be different. $\Delta\Delta G_{i,j}$ is the difference between the conservation energy for position $i$ ($\Delta G_i$) and a perturbed energy vector $\Delta G_{i,\delta j}$. The $\Delta G_i^x$ component of $\Delta G_i$ measures the probability of finding fragment $i.x$ relative to a reference probability:

$$\frac{\Delta G_i^x}{kT} = \frac{1}{N} \ln \frac{P_i^x}{P^*}. \quad [3]$$

$P_i^x$ is the binomial probability of fragment $i.x$ appearing $N_i^x$ times in a set of $N$ folded proteins. Assuming a reference state where all three fragments are equally likely,

$$P_i^x = \binom{N}{N_i^x} \frac{1}{3}^{N_i^x} \frac{2}{3}^{N-N_i^x}. \quad [4]$$

The reference probability $P^*$ follows Equation 4 with the substitution $N/3$ for $N_i^x$.

The three components of the perturbed energy are defined similarly to Equation 3:

$$\frac{\Delta G^x_{i,\delta j}}{kT} = \frac{1}{N} \ln \frac{P^x_{i,\delta j}}{P^*_{\delta j}}, \quad [5]$$

except now $N$ is the number of sequences in a subalignment containing only those folded sequences with fragment $j.1$. (Equivalently, one could perturb with respect to fragment 2 or fragment 3.) The binomial probabilities $P^x_{i,\delta j}$ and $P^*_{\delta j}$ follow Equation 4 with the appropriate parameters from the subalignment.

Equations 3 and 5 differ from those presented by Ranganathan and co-workers[2,21] in that the statistical energy vectors are normalized by alignment size. Without this normalization, the statistical energies scale linearly with alignment size, which is unphysical. Moreover, because the perturbation energy is calculated using only a subset of the alignment used for the conservation energy, the latter dominates the coupling energy when unnormalized. We observed this empirically when first implementing the published version of the algorithm.

**Excess information analysis**

The uncertainty about folding ($F$) in a set of $N$ sequences, only $N_f$ of which are folded, was quantified by the Shannon entropy.[25] If $p = N_f/N$ denotes the fraction folded, then the entropy in bits per sequence is

$$H(F) = -[p \log_2 p + (1-p)\log_2(1-p)] \geq 0. \quad [6]$$

Given no other information besides the ratio $p$, the likelihood of correctly predicting the folding status of every sequence is $2^{-NH}$. Systems with lower entropy are thus easier to predict. The conditional entropy $H(F|j.y)$, which must be less than or equal to $H(F)$, measures uncertainty when the presence or absence of fragment $j.y$ is known. It is defined by Equation 6 when $p$ is replaced with the conditional probability $p(F|j.y)$, which is the fraction of sequences with fragment $j.y$ that are also folded. The conditional entropy for position $j$ is the average over all three fragments:

$$H(F \mid j) = \sum_{y} p(j.y) H(F \mid j.y). \qquad [7]$$

The reduction in entropy $H(F) - H(F|j)$, which equals the mutual information $I(F:j)$ between folding and position $j$, represents how much the uncertainty about folding is reduced by knowing the fragment at position $j$.

The mutual information between folding and the position pair $i$-$j$ is defined similarly. Given that a sequence contains fragments $i.x$ and $j.y$, its probability of folding is $p(F|i.x, j.y)$. When substituted into Equation 6, this yields the conditional entropy $H(F|i.x, j.y)$. The conditional entropy for pair $i$-$j$ is the average over all nine fragment combinations:

$$H(F \mid i, j) = \sum_{x}\sum_{y} p(i.x, j.y) H(F \mid i.x, j.y). \qquad [8]$$

The mutual information between folding and pair $i$-$j$ is $I(F:i, j) = H(F) - H(F|i, j)$. The excess information for a pair, defined as the difference between the mutual information for the pair and the mutual information of its constituent positions, $I(F:i, j) - I(F:i) - I(F:j)$, was used to predict interactions.

**Logistic regression analysis**

Both logistic[18,19] and linear regression are special cases of the statistical methodology known as generalized linear modeling.[26,27] There are three components to a generalized linear model. The random component specifies a response variable $Y$ and its probability distribution. The systematic component specifies a predictor variable

$$\eta = \sum_k \beta_k \Gamma_k, \qquad [9]$$

which is a linear combination of explanatory variables ($\Gamma$). The use of a linear predictor variable does not preclude modeling interaction effects; each interaction is simply another explanatory variable. The third component of a generalized linear model is the link function $g(\cdot)$, which specifies the relationship between the mean of the response variable, $E[Y] = \mu$, and the predictor variable via $g(\mu) = \eta$. The choice of link function depends on the probability distribution of the response variable. Linear regression deals with normally distributed variables, for which the link is the identity function. In logistic regression, the response variable is binary and follows the Bernoulli (binomial) distribution, for which the logit function is the appropriate link:

$$\eta = \log\left(\frac{\mu}{1-\mu}\right). \qquad [10]$$

Inverting Equation 10 expresses the mean, which equals the probability of observing $Y = 1$, in terms of the predictor variable:

$$p(Y=1) = \mu = \frac{1}{1+e^\eta}. \qquad [11]$$

We used this framework to model whether a protein is folded ($F = 1$) or not ($F = 0$) in the combinatorial library. Each of the 3×8 = 24 fragments and 3×3×28 = 252 fragment-fragment pairs has a corresponding binary explanatory variable to model its presence ($\Gamma = 1$) or absence ($\Gamma = 0$). If we interpret the regression coefficients of these variables ($\beta$ in Equation 9) as one- and two-body energy terms, then Equation 11 models the probability of folding $p(F|E)$ as a sigmoidally decreasing function of the energy:

$$p(F|E) = \frac{1}{1+e^E}, \qquad \text{where} \qquad [12]$$

$$E = \varepsilon_0 + \sum_{i=1}^{8}\sum_{x=1}^{3}\varepsilon_1(i.x)\Gamma(i.x) + \sum_{i=1}^{7}\sum_{j=i+1}^{8}\sum_{x=1}^{3}\sum_{y=1}^{3}\varepsilon_2(i.x, j.y)\Gamma(i.x, j.y) \qquad [13]$$

The energy parameters in Equation 13 were fit by maximizing the likelihood function $L$, which is a product of $N$ terms, one for each chimera. If the $N_f$ folded chimeras are listed before the unfolded ones, then

$$L = \prod_{k=1}^{N_f} p(F|E_k) \prod_{k=N_f+1}^{N} [1 - p(F|E_k)]. \qquad [14]$$

Maximizing $L$ is equivalent to minimizing the deviance function,

$$D = -2 \ln L = -2 \sum_{k=N_f+1}^{N} E_k + 2\sum_{k=1}^{N} \ln(1 + e^{E_k}). \qquad [15]$$

Each $E_k$ in Equation 15 is a placeholder for an expression of the form given in Equation 13, in which the explanatory variables taken on the appropriate binary values for that sequence. This substitution leaves the energy terms as the only unknowns in Equation 15. Convex analysis[28] of Equation 15 reveals that the deviance is a convex function of the energy terms, which means a local optimization algorithm can be used to find globally optimal parameter estimates.

The significance of every position and every position pair was computed relative to a reference energy model, in which the two-body terms were constrained to zero while solving for the constant and one-body terms. Upon removing a position from this reference model, i.e., constraining its three one-body terms to equal zero, the minimum deviance must increase because there are fewer parameters to fit the data. The magnitude of this increase asymptotically follows the chi-square distribution with two degrees of freedom, which was used to compute a p-value for each position (i.e., we applied the likelihood ratio test). Conversely, adding a position pair by unconstraining its two-body terms lowers the minimum deviance, and the significance of this change was computed from the chi-square distribution with four degrees of freedom.

As just mentioned, each position has only two degrees of freedom despite the presence of three one-body terms. To understand this accounting, realize that any uniform shift in the one-body terms for a position can be counteracted by an equal and opposite shift in the constant term $\varepsilon_0$, thereby leaving the total energy unchanged. Only changes in a position's one-body terms relative to one another affect the deviance. We are thus free to set the average energy for each position equal to zero, which uniquely determines the one-body terms:

$$\sum_x \varepsilon_1(i.x) = 0. \qquad [16]$$

Similarly, in the presence of the one-body terms for a position pair, its nine two-body terms constitute only four degrees of freedom. To understand why, consider the two-body terms as a 3×3 matrix:

$$\chi \equiv \begin{bmatrix} \varepsilon_2(i.1, j.1) & \varepsilon_2(i.1, j.2) & \varepsilon_2(i.1, j.3) \\ \varepsilon_2(i.2, j.1) & \varepsilon_2(i.2, j.2) & \varepsilon_2(i.2, j.3) \\ \varepsilon_2(i.3, j.1) & \varepsilon_2(i.3, j.2) & \varepsilon_2(i.3, j.3) \end{bmatrix}. \qquad [17]$$

A uniform shift in row $x$ or column $y$ of this matrix can be counteracted by an equal and opposite shift in the one-body term $\varepsilon_1(i.x)$ or $\varepsilon_1(j.y)$, respectively, for which the degree of freedom has been previously counted. We are thus free to set each column and row average equal to zero, thereby uniquely determining the two-body terms:

$$\sum_x \varepsilon_2(i.x, j.y) = \sum_y \varepsilon_2(i.x, j.y) = 0. \qquad [18]$$

For each position pair *i-j*, Equation 18 represents six linear constraints, but it is straightforward to show that only five are linearly independent, e.g., by rewriting Equation 18 as the vector equation $\mathbf{A}\boldsymbol{\chi} = \mathbf{0}$ and computing the rank of the 6×3 matrix $\mathbf{A}$. We conclude that nine two-body terms minus five linearly independent constraints equals four degrees of freedom.

To summarize, our logistic regression algorithm involves minimizing the deviance in Equation 15 subject to the 8 linear constraints in Equation 16 and the (3 rows + 3 columns)×(28 position pairs) = 168 linear constraints represented by Equation 18. Furthermore, some energy terms may be set equal to zero as described in the paragraph on establishing statistical significance. Since these constraints are all linear, the optimization problem remains convex,[28] which greatly simplifies the task of fitting the energy terms. After encoding this problem with the modeling language GAMS, we submitted jobs to the NEOS server for optimization.[29] The algorithm MINOS worked well, with execution times typically less than one second.

## Results

The folding status of every sequence in the fictitious library was determined according to the hypothetical energy model summarized in Figure 2A. The diagonal entries of this 8×8 matrix represent the individual, or one-body, contributions of the eight positions, and the off-diagonal entries represent the interaction, or two-body, strengths of the position pairs. The energy model (see Tables I and II) was designed with two complementary challenges in mind. Algorithms should detect two-body interactions when they are present, but at the same time interactions should not be predicted when none exist. Both tasks are present in Figure 2A, where for simplicity only a single pair of positions (2-7) was chosen to interact. Except for position 2, whose one-body terms were set to zero, the one-body terms were chosen randomly to give a spectrum of contributions to stability. In order of decreasing strength, the positions are 4, 5, 7, 8, 3, 1, 6, and 2.

Panels B through E in Figure 2 show the predictions of four different algorithms. Except for contingency table analysis (panel B), which does not score the one-body terms, the algorithms make qualitatively correct predictions about the relative importance of the individual contributions made by the positions. In addition, all but statistical coupling analysis (panel C) satisfactorily detect the interaction between positions 2 and 7.

However, only logistic regression (panel E) avoids making spurious two-body predictions. The contingency table (panel B), statistical coupling (panel C), and excess information (panel D) algorithms all score pair 4-5 as having the strongest interaction. These positions do not interact in the energy model, but they strongly affect folding as individual fragments (see Figure 2A). When present in the same sequence, stabilizing fragments tend to lower the energy below the threshold needed for folding. As a result, their frequencies in the alignment can be correlated even if the positions do not interact. In panels B, C, and D of Figure 2, it is clear that the stronger the individual contribution of a position, the more frequently it is predicted to make spurious interactions. Statistical coupling analysis appears most susceptible to this error, followed by the contingency table and excess information algorithms.

## Discussion

The ability of logistic regression to avoid spurious predictions stems not only from its use of the unfolded sequences, a characteristic it shares with the excess information analysis depicted in Figure 2D. Its distinguishing feature is the ability to directly test for energetic coupling with a sigmoidal folding model (Equation 12). The other three algorithms are only able to test for probabilistic coupling, i.e., whether the probability distributions for two positions are independent. This is not a reliable indicator of energetic coupling when folding is a nonlinear function of energy.

Are these results relevant to real proteins? We have applied logistic regression to a combinatorial library of cytochrome P450 proteins and found the predictions to be consistent with structural and dynamic studies of that protein family.[15] The energetic patterns observed in the P450 data resemble our hypothetical energy model in that both have a single dominant two-body term and a spectrum of one-body strengths. For proteins with more limited structural information, logistic regression could help test putative interactions or otherwise refine a homology model. Because the sigmoid used by logistic regression to link sequence and function (Equation 12) is quite general, other properties besides protein folding can be modeled. We are particularly interested in probing the sequence determinants of enzyme specificity. High-throughput screening for catalytic activity on a particular substrate can generate the binary data needed for this analysis.

As with any multi-parameter estimation, one of the difficulties encountered when applying logistic regression to real data is establishing statistical significance. As explained in the methods, the entropy of folding in a data set increases with the number of sequences and as the fraction folded approaches ½ Each time an energy term is estimated, the entropy of the data set decreases, and hence the initial entropy limits how many parameters can be reliably estimated. (Nonuniform patterns of sampling from the library influence which particular energy terms can be fit.) Our fictitious library had high entropy because all of the sequences were characterized and close to half were folded, but this will not be true of real combinatorial libraries (e.g., our cytochrome P450 data set in which only 806 of the 6,561 sequences were

characterized[15]) and certainly does not hold for MSAs of natural proteins. Natural MSAs contain very few of the total possible sequences, all of which are folded and functional (zero entropy); this precludes the use of logistic regression.

The weaknesses we have observed with contingency table and statistical coupling analysis on the fictitious library seem relevant to their performance on natural MSAs. Both algorithms can be duped into predicting spurious interactions between relatively conserved positions, despite being formulated to avoid this ubiquitous pitfall. In hindsight it seems obvious that any linear statistical analysis will make this error if folding depends nonlinearly on free energy. This may partially explain why, for the majority of protein families, fewer than 30% of predicted interactions involve contacting residues, even with the best algorithms.[5,6,30,31] (Of course not all interactions detectable in an MSA are short-ranged, so this measure of accuracy is necessarily pessimistic.)

## Acknowledgments

We thank Yaser Abu-Mostafa, Allan Drummond, Claus Wilke, and two anonymous referees for their valuable suggestions. This work was supported by the National Institutes of Health (grant R01 GM068664-01), the Institute for Collaborative Biotechnologies (Army Research Office grant DAAD19-03-D-0004), a National Defense Science and Engineering Graduate Fellowship (J.B.E.), and a Howard Hughes Medical Institute Pre-Doctoral Fellowship (J.D.B.).

**Table I.** One-body terms in the hypothetical energy model.

| Position | Fragment 1 | Fragment 2 | Fragment 3 |
|---|---|---|---|
| 1 | 0.56 | 0.21 | -0.77 |
| 2 | 0 | 0 | 0 |
| 3 | 0.83 | -1.00 | 0.17 |
| 4 | -2.24 | -0.45 | 2.69 |
| 5 | 1.10 | 1.40 | -2.50 |
| 6 | -0.00 | -0.46 | 0.46 |
| 7 | -1.61 | 0.03 | 1.58 |
| 8 | -1.23 | 0.80 | 0.43 |

**Table II.** Two-body terms for pair 2-7 in the hypothetical energy model. All other two-body terms are zero. Fragment *i.x* refers to fragment *x* at position *i*.

|     | 7.1   | 7.2   | 7.3   |
|-----|-------|-------|-------|
| **2.1** | 0.46  | 0.38  | -0.84 |
| **2.2** | 0.63  | -1.01 | 0.38  |
| **2.3** | -1.09 | 0.63  | 0.46  |

**Figure legends**

**Fig. 1.** Combinatorial protein libraries of known size and composition. **(A)** In the class of libraries considered here, amino acid fragments (of any length, including one residue) are incorporated at a fixed number of variable positions. In this example, each of the eight variable positions (shown as open blocks) has three possible fragments, for a total of $3^8 = 6,561$ sequences. **(B)** In site-directed recombination, the fragments are taken from homologous sequences (denoted parents 1, 2, and 3) and designed to completely span the protein. **(C)** Sequences from the libraries in panels A and B are aligned the same way, with each variable position represented as a single column regardless of fragment length. Instead of amino acid letters, the sequences contain fragment numbers.

**Fig. 2.** Logistic regression is the only algorithm to correctly predict pairwise interactions in a library of fictitious proteins. **(A) Energy model.** This symmetric 8×8 matrix summarizes the hypothetical energy model used to fold the fictitious proteins (Tables I and II). The diagonal entries reflect the individual contribution of each position, and the off-diagonal entries represent pairwise interactions. The shade of each square encodes the standard deviation across energy parameters, which is a measure of energetic "strength." Because the average energy for each variable is arbitrarily constrained to zero, the standard deviation is $\sqrt{\frac{1}{3}\sum_{x}[\varepsilon_1(i.x)]^2}$ for position $i$ and $\sqrt{\frac{1}{9}\sum_{x}\sum_{y}[\varepsilon_2(i.x, j.y)]^2}$ for pair $i$-$j$. **(B) Contingency table analysis.** The shade of each off-diagonal entry encodes the chi-square statistic with four degrees of freedom. Larger values indicate the fragments are not distributed independently, but this does not mean they interact energetically. This is clear from a comparison with the energy model in panel A. The diagonal

entries are blank because contingency table analysis does not score one-body terms.

**(C) Statistical coupling analysis.** Conservation energies are shown on the diagonal and coupling energies are shown off the diagonal. The shade of each square represents the magnitude of the corresponding statistical energy vector, in units where $kT = 100$. Although the one-body effects are predicted well, the two-body interactions are not. **(D) Excess information analysis**. The shade of each diagonal entry represents the mutual information with folding for that position, in bits per sequence. The off-diagonal entries reflect the excess information for each pair in bits per sequence. Although this algorithm uses both folded and unfolded sequences, there is little improvement over the contingency table analysis in panel B.

**(E) Logistic regression analysis.** The shade of each square represents the significance p-value from the likelihood ratio test, reported as $-\log_{10}(p)$ so that higher numbers are more significant. Panels A and E are nearly indistinguishable, which means logistic regression is able to determine all the important energy terms and their relative strengths.

# Figure 1

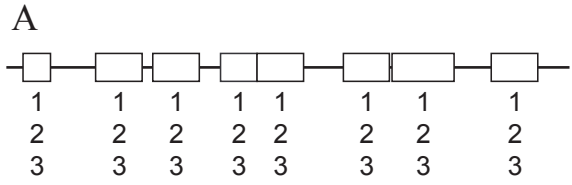

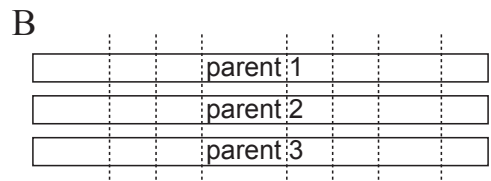

C
```
12311223
11113321
12121331
33231233
    ⋮
```

# Figure 2

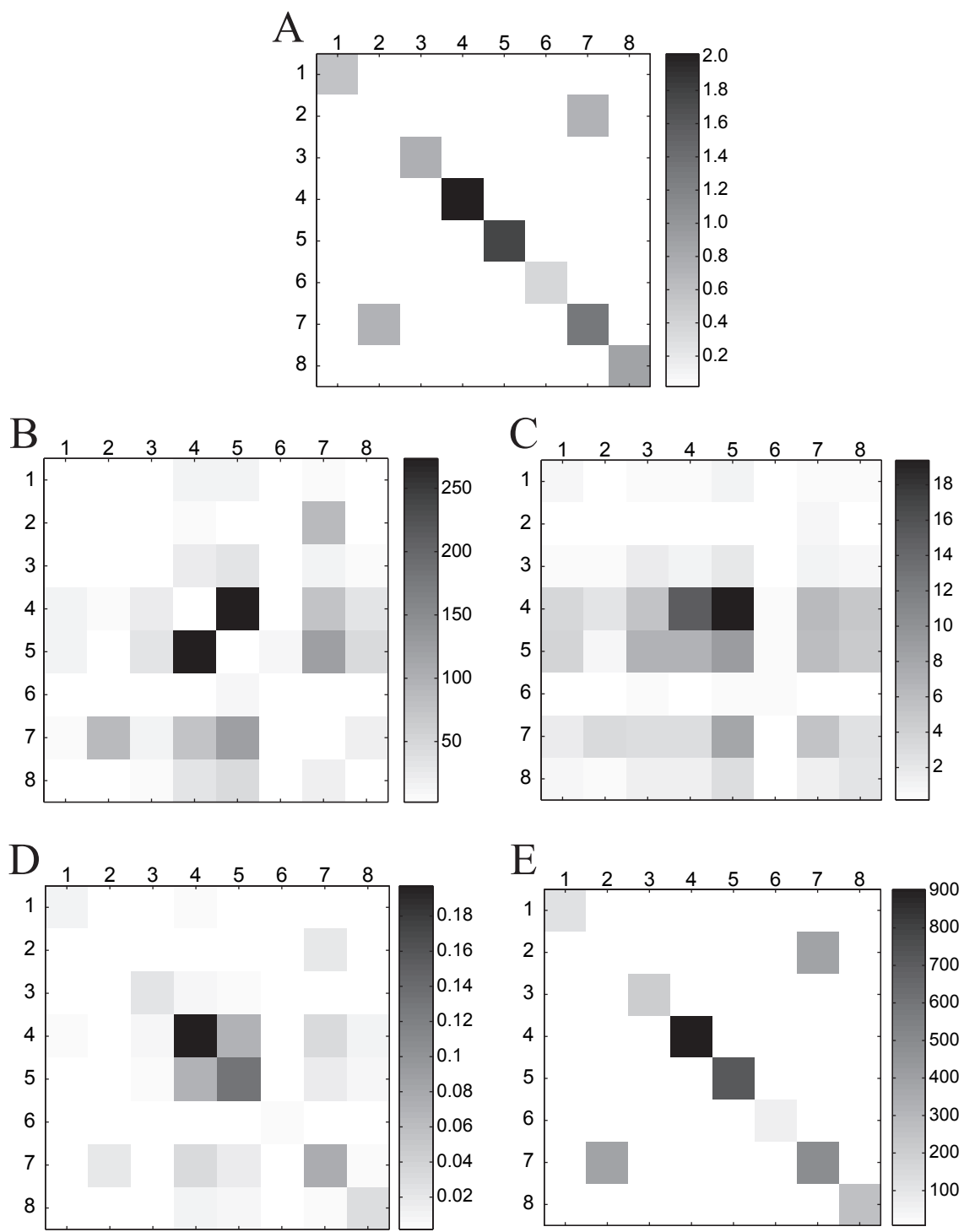